\documentclass[iop]{emulateapj}
\usepackage{natbib}
\usepackage{txfonts}
\usepackage{graphicx}
\usepackage{wasysym}
\usepackage{color}

\newcommand{\ha}{H$\alpha$}

\newcommand{\fluxunit}{ergs s$^{-1}$ cm$^{-2}$}
\newcommand{\sbunit}{ergs s$^{-1}$ cm$^{-2}$ arcsec$^{-2}$}

\newcommand{\hal}{L$_{\rm H\alpha}$}
\newcommand{\NII}{[\ion{N}{2}]}
\newcommand{\SII}{[\ion{S}{2}]}
\newcommand{\OI}{[\ion{O}{1}]}

\begin{document}

\title{Discovery of a vast ionized gas cloud in the M51 system}

\shortauthors{Watkins et al.}
\shorttitle{M51 H$\alpha$ cloud}

\author{Aaron E. Watkins\altaffilmark{1,2},
  J. Christopher Mihos\altaffilmark{2},
  Matthew Bershady\altaffilmark{3},
  Paul Harding\altaffilmark{2}}

\altaffiltext{1}{Astronomy Research Unit, University of Oulu,
  FIN-90014, Finland}
\altaffiltext{2}{Department of Astronomy, Case Western Reserve
  University, Cleveland, OH 44106, USA}  
\altaffiltext{3}{Department of Astronomy, University of Wisconsin, 475
  N. Charter Street, Madison, WI 53706, USA}

\begin{abstract}

We present the discovery of a vast cloud of ionized gas 13\arcmin\ (32
kpc) north of the interacting system M51. We detected this cloud via
deep narrow-band imaging with the Burrell Schmidt Telescope, where it
appears as an extended, diffuse \ha-emitting feature with no embedded
compact regions. The Cloud spans
$\sim$10\arcmin$\times$3\arcmin\ (25$\times$7.5 kpc) in size and has
no stellar counterpart; comparisons with our previous deep broadband
imaging show no detected continuum light to a limit of $\mu_{\rm lim,
  B} \sim$30 mag arcsec$^{-2}$. WIYN\footnote{The WIYN
    Observatory is a joint facility of the University of
    Wisconsin-Madison, Indiana University, the National Optical
    Astronomy Observatory and the University of Missouri.}\ SparsePak
  observations confirm the cloud's kinematic association with M51,
and the high \NII$/$\ha, \SII$/$\ha, and \OI$/$\ha\ line ratios we
measure imply a hard ionization source such as AGN photoionization or
shock heating rather than photoionization due to young stars. Given
the strong \NII\ emission, we infer roughly solar metallicity for the
cloud, ruling out an origin due to infall of primordial gas. Instead
we favor models where the gas has been expelled from the inner regions
of the M51 system due to tidal stripping or starburst/AGN winds and
has been subsequently ionized either by shocks or a fading AGN. This
latter scenario raises the intriguing possibility that M51 may be the
nearest example of an AGN fossil nebula or light echo, akin to
the famous ``Hanny's Voorwerp" in the IC~2497 system.

\end{abstract}


\newpage

\section{Introduction}

The galaxy pair M51 (NGC~5194/5) is perhaps the most iconic
interacting system. It has been a subject of study since the 1800s,
its spiral structure serving as a key element in early debates over
the true nature of galaxies \citep{steinicke12}.  It has served, via
simulations and observations, as an important dynamical benchmark for
studies of tidal tails \citep[e.g.,][]{toomre72, burkhead78, rots90,
  salo00a} and spiral density waves \citep[e.g.,][]{dobbs10}. Because
of its active star formation \citep[star formation rate (SFR) $\sim
  1.6$ M$_{\odot}$ yr$^{-1}$;][]{kennicutt09} and close distance
\citep[8.6 Mpc;][]{mcquinn16}, it is also often used to calibrate star
formation tracers and gas density--SFR relations
\citep[e.g.,][]{calzetti05, kennicutt07}. Yet despite this scrutiny,
the system continues to yield surprises.

For example, while M51 is known to host extended gaseous tidal debris
\citep{rots90}, the detailed impact of the interaction and subsequent
star formation and nuclear activity on this extended gas is poorly
understood.  Narrow-band imaging by \citet{hoopes01} also shows an
anomalous ``hook'' of ionized gas overlapping the companion galaxy,
possibly stripped from the primary and shock-heated during the
interaction.  However, while M51 has been the frequent target of
narrow-band imaging and spectroscopy \citep[e.g.,][]{hoopes01,
  thilker02, calzetti05, kennicutt07}, to date these have lacked the
wide field coverage necessary to explore the pair's extreme outskirts
where more extended and diffuse ionized gas could reside.

In this Letter, we present the discovery of a new such feature in M51:
a large, diffuse circumgalactic cloud of ionized gas (hereafter, the
Cloud), spanning $\sim$10\arcmin$\times$3\arcmin\ (25$\times$7.5 kpc)
in size and located $\sim$13\arcmin\ (32 kpc) north of NGC~5194's
center.  We discovered this Cloud through deep wide-field narrow-band
imaging using the Burrell Schmidt Telescope at Kitt Peak National
Observatory (KPNO).  We present the results of this imaging, as well
as follow-up spectroscopy done with the Sparsepak Integral Field
  Unit (IFU) at the WIYN 3.5m Telescope \citep{bershady04,
  bershady05}.  We discuss the Cloud's possible origins given its
properties and our current understanding of the M51 system.

\section{Observations and Data Reduction}

\subsection{Narrow-band imaging}

\begin{figure*}
  \centering
  \includegraphics[scale=0.61]{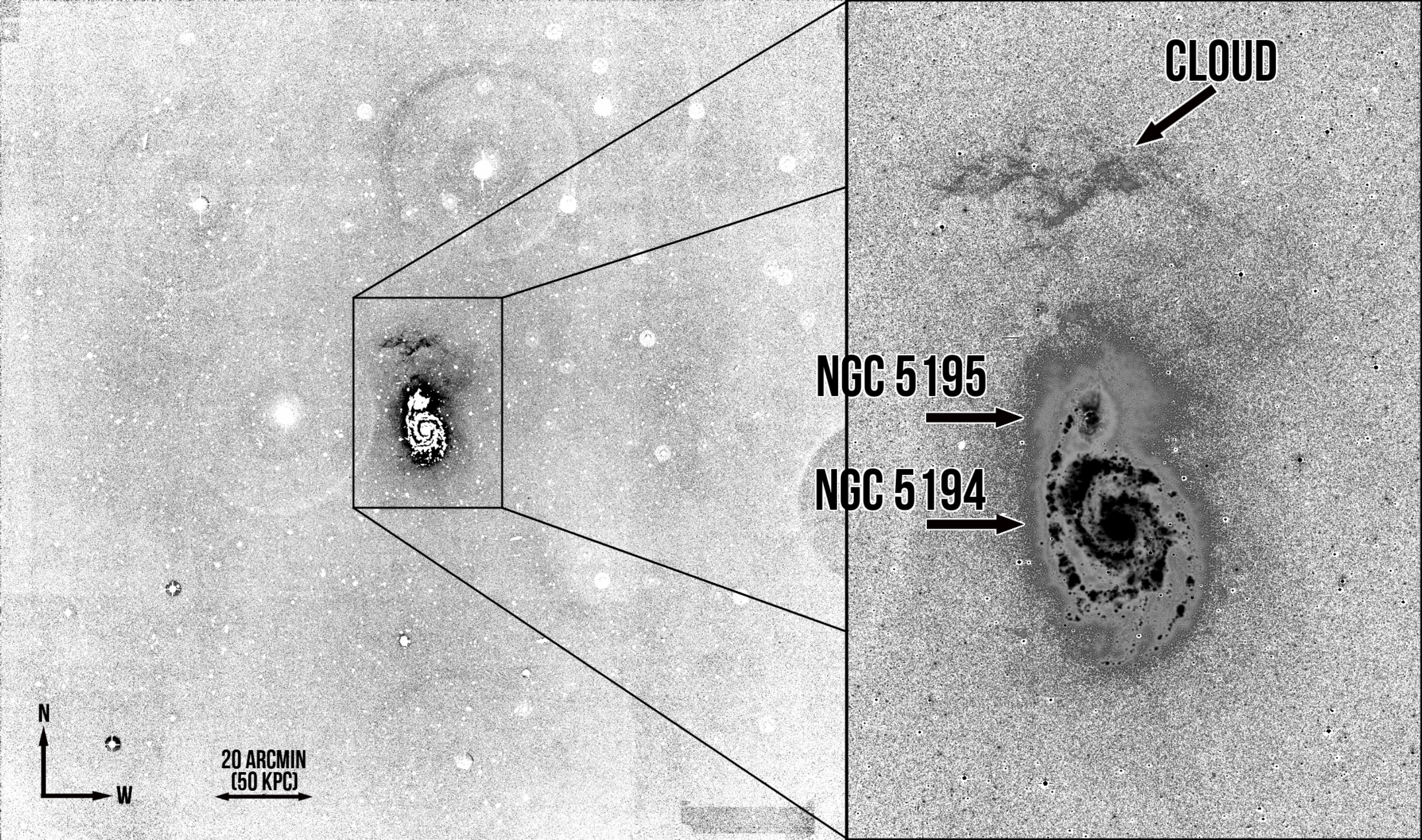}
  \caption{{\bf Left:} \ha\ difference mosaic (see text), masked of
    bright sources and 3$\times$3 pixel median-binned to emphasize
    diffuse \ha\ emission in M51 and the surrounding environment.
    The image has a per-pixel RMS noise of
      5.7$\times$10$^{-19}$\sbunit; the faintest emission visible is
      at $\sim$10$^{-18}$ \sbunit.  {\bf Right:} a zoomed-in view,
    showing the unbinned \ha\ difference mosaic at high dynamic range.
    \label{fig:mosaic}}
\end{figure*}

We observed M51 with the Burrell Schmidt in the Spring of 2015 using
two custom $\sim$100 \AA-wide narrowband filters designed to target
\ha$+$\NII\ emission and the adjacent continuum in nearby (D$<20$ Mpc)
galaxies. We observed M51 only on moonless, photometric nights in
March, April, and May. The Schmidt field of view covers
  1\fdg65$\times$1\fdg65; in each filter we imaged M51 in
  77$\times$1200s images, randomly dithering pointings by
  $\sim$30\arcmin\ around the target. We also observed 33 offset
night sky fields alongside $\sim$100 twilight exposures to build flat
fields \citep[described further below; however, for details,
  see][]{watkins17}.  Finally, we targeted a variety of
spectrophotometric standard stars from \citet{massey88} for flux
calibration.

We began data reduction with standard bias subtraction and nonlinearity
corrections. For flat-fielding, we used a combination of night sky and
twilight sky images. Due to the low count rate in the narrow-band night
sky images, we combined the twilight sky images to build preliminary
flat fields. However, because of the Burrell Schmidt's large field of
view, these flats contain strong gradients induced by the twilight sky; we
thus used the night-sky flats to model and remove these linear gradients
\citep{watkins17}. To remove mild ($<$1\% amplitude) fringing from the
on-band exposures, we isolated the fringe pattern from the combined
on-band night-sky flats through division by the on-band twilight flat
(where the fringe pattern was not visible against the much brighter flat
field), then scaled and subtracted a normalized version of this fringe
map from each frame using IRAF's\footnote{IRAF is distributed by the
National Optical Astronomy Observatory, which is operated by the
Association of Universities for Research in Astronomy (AURA) , Inc.,
under cooperative agreement with the National Science Foundation.}
RMFRINGE package.

To reduce the effects of scattered light, we used long (1200s)
exposures of Arcturus to model and remove internal reflections and the
extended wings of the PSF from bright ($V \lesssim$10 mag) stars in
each frame \citep[see][]{slater09}.  For each frame, we then
hand-masked all stars, galaxies, and any scattered light artifacts,
modeled the remaining sky as a plane, and subtracted it.  Finally, we
median combined the images to create three final image stacks
  with total exposure times of 25.6 hr: an on-band mosaic (composed
of all on-band exposures), an off-band mosaic (similarly defined), and
a ``difference'' mosaic created by combining individual difference
images between on-off pairs observed sequentially in time (Figure
\ref{fig:mosaic}).  We use the difference images to illustrate
  the morphology of the diffuse \ha, but conduct quantitative
  photometry on the on- and off-band mosaics directly.

\subsection{Sparsepak spectroscopy}

\begin{figure*}
  \centering
  \includegraphics[scale=0.5]{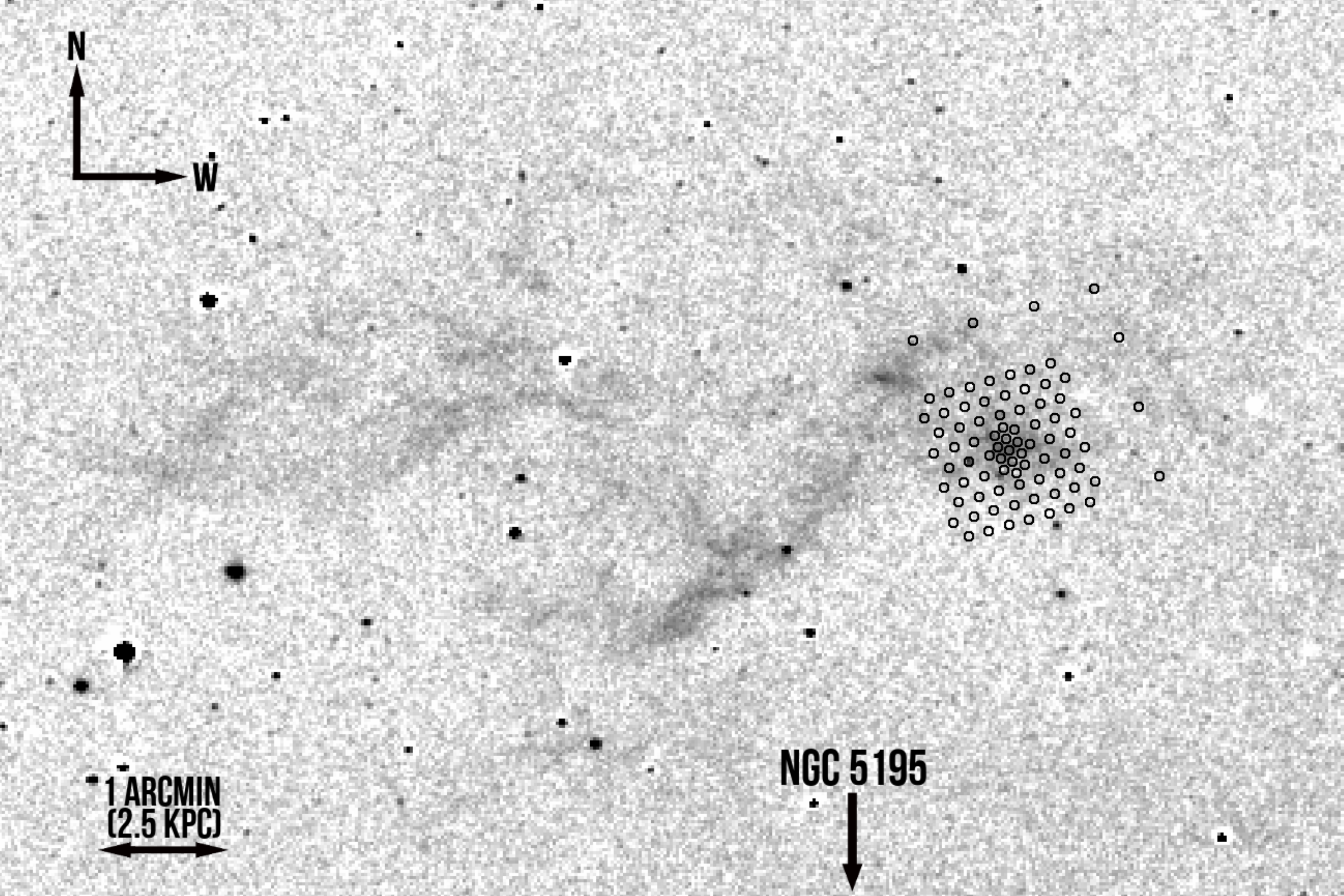}
  \caption{Close-up of the Cloud in the \ha\ difference image, with
    the Sparsepak fiber array orientation overlaid.  
    \label{fig:sparsepak}}
\end{figure*}

We obtained follow-up spectroscopy of the Cloud in April 2017 using
Sparsepak \citep{bershady04, bershady05}.  Sparsepak, comprised of 82
4.687\arcsec\ diameter fibers, feeds into the WIYN Bench Spectrograph.
We centered Sparsepak on the region with the brightest emission, where
the 72\arcsec~$\times$~72\arcsec coverage spans the far western end of
the Cloud (Figure \ref{fig:sparsepak}).  To target \ha\ emission, we
used the 860 line mm$^{-1}$ grating in 2$^{\rm nd}$ order, to cover
6187\AA\ to 7096\AA\ at a resolution of $\lambda / \delta\lambda =$
3380 ($\sigma \sim$37 km s$^{-1}$).  We fully resolve \ha,
\NII\ 6548/6583\AA, and \SII\ 6716/6731\AA\ from each target.

We took 1200s exposures of the target, chopping to a nearby sky field
on occasion to yield more accurate sky subtraction.  Our final object
spectrum contains 31 target exposures, with 23 accompanying sky
exposures of the same length.  We used the IRAF tasks IMCOMBINE and
CCDPROC for basic image processing (bias, overscan and dark
subtraction, and cosmic ray removal) and DOHYDRA for spectral
extraction, channel-to-channel correction (via accompanying dome flat
exposures), and wavelength calibration (via accompanying ThAr arc lamp
exposures).  Calibration data were taken over the course of the run.

\begin{figure*}
  \centering
  \includegraphics[scale=0.8]{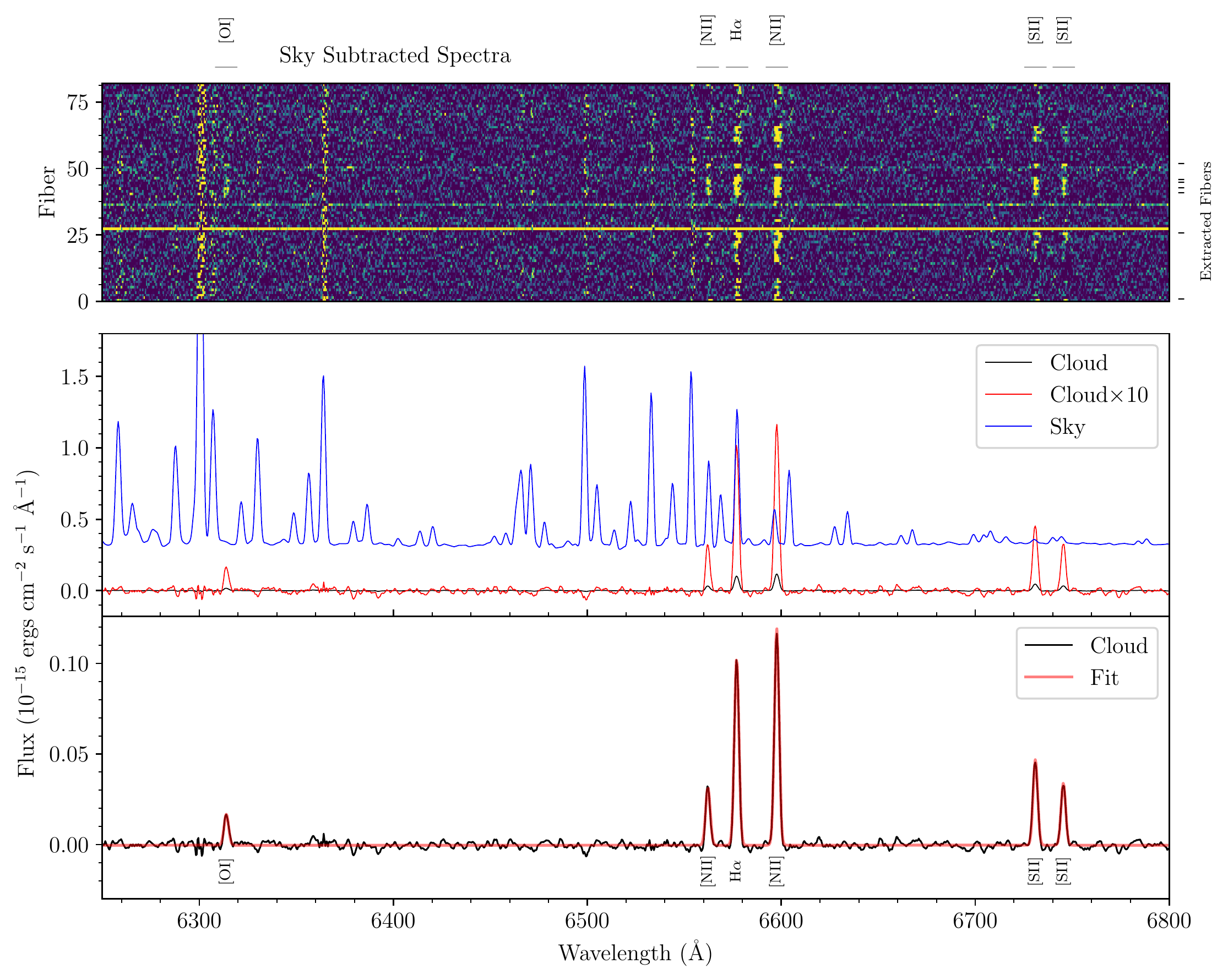}
  \caption{{\bf Top:} final reduced extracted 2D spectrum of the
    Cloud.  The seven fibers used in co-addition are marked on the
    right. {\bf Middle:} final reduced co-added spectrum of the Cloud.
    The reduced spectrum is shown in black; the red curve scales this
    spectrum's intensity up by $\times$10 to show weak emission lines.
    The foreground sky spectrum (blue) is shown to illustrate the
    strong overlap between the Cloud's redshifted emission lines and
    telluric lines.  {\bf Bottom:} fitted spectrum (red) overlaid on
    the Cloud's reduced spectrum (black).
    \label{fig:spectra}}
\end{figure*}

Due to the target's low surface brightness, we subtract sky
independently of DOHYDRA, augmenting the method described in Appendix
D of \citep{bershady05} to take advantage of independent sky
exposures. Briefly, we first we fit low-order polynomials in
wavelength to each extracted fiber aperture in the combined sky
frames, rejecting sky-lines via clipping, to produce a sky flat. The
flat is normalized and applied to both combined object and sky
extracted spectra. The sky spectra are then subtracted from the object
spectra, fiber by fiber; this yields excellent subtraction of the sky
continuum but leaves residuals in the sky-lines due to different
line-to-line temporal variations. To remove these sky-line residuals
we fit low-order polynomials, with clipping, in the pseudo-slit
dimension for each wavelength channel. The resulting reduced spectrum
is shown in \ref{fig:spectra}, coadded from the seven fibers with the
strongest [N~II]$\lambda$6583\AA\ signal. The sky spectrum is shown for
comparison; note that at this Cloud's velocity, all five targeted
lines strongly overlap with telluric line emission.

\begin{deluxetable}{l l}
  \tabletypesize{\small}
  \tablecaption{Cloud Properties\label{tab:prop}}
  \tablecolumns{2}
  \tablehead{
    }
    \startdata
    \\[0.01cm]
    ($\alpha$, $\delta$)$_{\rm peak}$  &  13h29m51.1s, $+$47d24m34s  \\[0.01cm]
    Dimensions  &  10\arcmin$\times$3\arcmin \ (25$\times$7.5  kpc)  \\[0.01cm]
    f$_{\rm H\alpha, tot}$  &  1.8$\times$10$^{-13}$   \fluxunit  \\[0.01cm]
    L$_{\rm H\alpha, tot}$  &  1.6$\times$10$^{39}$   erg s$^{-1}$  \\[0.01cm]
    Peak $\Sigma_{\rm H\alpha}$  &  2.2$\times$10$^{-16}$   \sbunit  \\[0.01cm]
    V$_{\rm heliocentric}$  &  637 $\pm$ 13   km s$^{-1}$  \\[0.01cm]
    $\log$([\ion{N}{2}]6748,6583~$/$~\ha)  &  $+$0.17  \\[0.01cm]
    $\log$([\ion{N}{2}]6583~$/$~\ha)  &  $+$0.06  \\[0.01cm]
    $\log$([\ion{S}{2}]6717,6731~$/$~\ha)  &  $-$0.10  \\[0.01cm]
    $\log$([\ion{S}{2}]6717~$/$~[\ion{S}{2}]6731)  &  $+$0.14   \\[0.01cm]
    $\log$([\ion{O}{1}]6300~$/$~\ha)  &  $-$0.77   \\[0.01cm]
    $\log$([\ion{O}{3}]5007~$/$~\ha)  &  $< -$0.64
    \enddata
\end{deluxetable}

\section{Results}

We provide the Cloud's basic properties in Table \ref{tab:prop}, while
Figure \ref{fig:mosaic} shows the Cloud's morphology and position
relative to M51. It is expansive ($\sim$25 kpc long), and though it
is near M51 on the sky, aside from some very faint ($\Sigma_{\rm
  H\alpha} \lesssim$1.6 $\times$ 10$^{-18}$ \sbunit) diffuse emission
northwest of the companion it shows no clear connection to the system.
From polygonal aperture photometry \citep[e.g.,][]{watkins15} of the
narrow-band imaging, we measure the Cloud's total
\ha$+$\NII\ luminosity as $\sim$4 $\times$10$^{39}$ ergs
s$^{-1}$. From the co-added Sparsepak spectra, the Cloud's mean
\NII\ (6583$+$6548)$/$\ha \ ratio is $\sim$1.48, hence the Cloud's
total \ha\ luminosity is \hal$\sim$1.6$\times$10$^{39}$ ergs s$^{-1}$.

We detect this \ha \ emission, the \NII\ and \SII\ doublets, and faint
\OI$\lambda$6300\AA \ in seven of the Sparsepak fibers; \ha \ and
\NII$\lambda$6583\AA \ are visible in an additional 17 fibers. Doing a
combined fit to the \NII, \SII, and \ha\ lines for the seven fibers
with the brightest \NII\ emission, we measure a heliocentric velocity
of 637 $\pm$ 13 km s$^{-1}$; the quoted uncertainty is the standard
deviation of the individual fiber measurements, most of which is
astrophysical variance in radial velocity. This velocity is well
outside the realm of most Milky Way emission, including high velocity
clouds (HVCs), which typically have velocities $<$ 500 km s$^{-1}$
\citep[and none of which have been discovered so near M51 on the
  sky;][]{westmeier18}. The Cloud's observed velocity is within the
range of \ion{H}{1} velocities observed in the M51 system
\citep{rots90}, confirming its kinematic association with the
interacting pair.

\begin{figure*}
  \centering
  \includegraphics[scale=0.43]{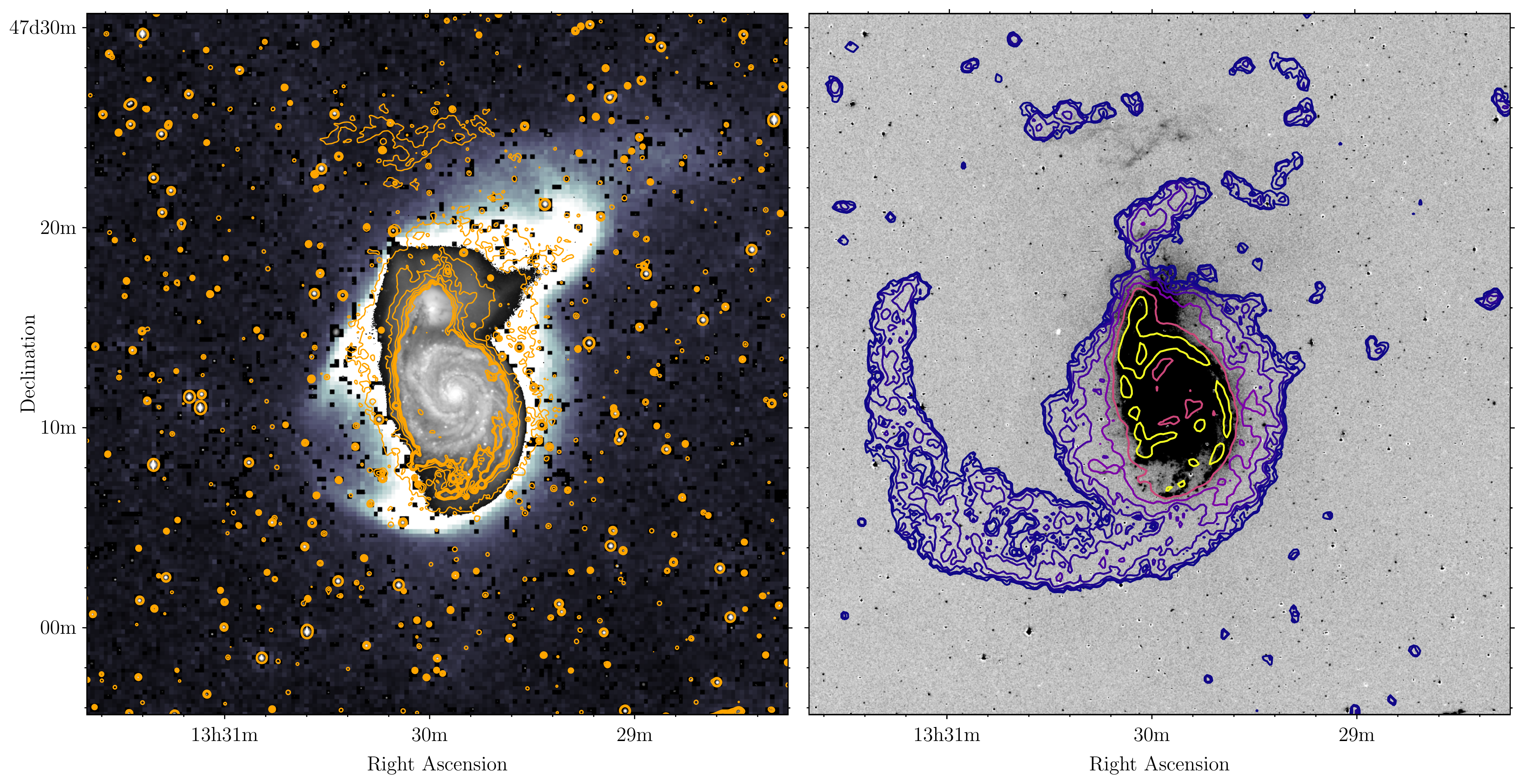}
  \caption{{\bf Left:} \ha-emission contours (orange) overlaid on a
    deep \emph{B}-band image of M51 from \citet{watkins15}.  The inner
    regions of M51 have been rescaled in intensity to show its high
    surface brightness structure.  {\bf Right:} \ion{H}{1} contours
    from the 34\arcsec\ resolution VLA interferometry map of
    \citet{rots90} overlaid on our \ha\ difference mosaic.
    \label{fig:comps}}
\end{figure*}

In the left panel of Figure \ref{fig:comps}, we overlay contours of
\ha \ emission (convolved with a 3$\times$3 pixel Gaussian kernel to
emphasize diffuse features) on a deep Burrell Schmidt \emph{B}-band
image of M51 \citep{watkins15}. The cloud appears to have no stellar
counterpart; it is undetected in the broadband imaging to a limit of
$\mu_{B} \sim$30 mags arcsec$^{-2}$, nor is it detected in our
similarly deep Washington \emph{M} image \citep{watkins15}.  The three
apertures with spectral continuum in Figure \ref{fig:spectra} each
have line-emission consistent with background galaxies.  Its
non-detection in Washington \emph{M} also implies a low
[\ion{O}{3}]$\lambda$5007\AA\ flux: $< 5\times10^{-17}$ \sbunit,
  or $\log($[\ion{O}{3}]/\ha$)< -0.64$. The
Cloud lies just east of M51's diffuse northwestern stellar plume, and
shows morphological contiguity with the northernmost of the bifurcated
western streams extending from the companion, whose faintest extension
arcs toward it.  We also note a marginal detection of diffuse far
ultraviolet emission near the Cloud \citep{bigiel10}, however it is
coincident with 12 $\mu$m emission visible in the {\it WISE} Galactic
cirrus map \citep{meisner14} of the region, and hence cannot be firmly
associated with the Cloud.

The right panel of Figure \ref{fig:comps} shows \ion{H}{1} contours
from the interferometry map of \citet{rots90} overlaid on our
\ha\ difference mosaic. While no high column density \ion{H}{1}
($>$3.3$\times$10$^{19}$ cm$^{-2}$) lies directly coincident with the
Cloud, diffuse \ion{H}{1} potentially associated with it is found
throughout the region (Pisano, private communication). We measure a
$\sim40$ km s$^{-1}$ gradient across the central fiber bundle,
oriented northeast to southwest (with the highest velocities in the
east); this orientation is similar to velocity gradients across the
two high velocity \citep[520--660 km s$^{-1}$;][]{rots90} \ion{H}{1}
clouds straddling it, implying a potential connection as well.

The \NII, \SII, and \OI\ lines give additional information on the
Cloud's ionization source.  From the co-added spectrum (Figure
\ref{fig:spectra}), we measure $\log$(\NII$\lambda$6584/\ha) $=
+0.06$, $\log$(\SII$\lambda\lambda$6716,31/\ha) $= -0.10$, and
$\log$(\OI$\lambda$6300/\ha) $= -0.77$. Variations in the sky
subtraction model yield line ratio uncertainties of $<0.1$ dex,
  similar to fiber-to-fiber differences between individual fiber
  spectra across the field.  These line ratios are significantly
higher than those in \ion{H}{2} regions, and are more typical of hard
photoionization from AGN \citep[e.g.,][]{kewley06}.  In AGN-ionized
clouds, the \NII/\ha\ line ratio is largely a metallicity indicator
\citep[e.g.,][]{fu09} over a wide range of ionization parameters, and
in the case of the M51 Cloud, suggests roughly solar metallicity for
the gas.

Alternatively, the gas may be shocked.  From MAPPINGS III
\citep{allen08} shock models, the Cloud's \NII$/$\ha, \SII$/$\ha, and
\OI$/$\ha\ line ratios are well-matched by shock velocities between
$\sim$200--300 km s$^{-1}$.  In these shock models as well, \NII$/$\ha
\ favors solar metallicity or higher.

In summary, the Cloud's relatively high velocity rules out a chance
projection of foreground Milky Way gas, and demonstrates the Cloud's
kinematic association with tidally stripped \ion{H}{1} gas in the M51
system. The lack of broadband light and the high \NII/\ha, \SII/\ha,
and \OI/\ha\ line ratios argue against ionization from young stars and
instead suggest ionization from AGNs or shocks. With the Cloud's
association with M51 now well-established, in the next section we
explore possible scenarios to explain the Cloud's origin and anomalous
line ratios.

\section{Discussion}

While the Cloud's physical association with M51 is well-demonstrated,
its origin is less clear. M51 is dynamically complex, with active star
formation \citep{kennicutt09}, numerous and disorganized tidal
features \citep{rots90, watkins15}, and nuclear activity in both
galaxies \citep{ho97, rampadarath18}.  In such a system, a number of
plausible scenarios are available to explain the Cloud's origin,
including tidal or ram-pressure stripping during the interaction,
infall from the surrounding circumgalactic environment, or ejection
from a central starburst or AGN.  Constraints on the metallicity and
excitation mechanism at work in the Cloud can help differentiate
between these various possibilities.

The roughly solar metallicity indicated by the \NII/\ha\ line ratio
constrains the gas' provenance. Such a high metallicity argues
  against scenarios that involve gas infalling from the
circumgalactic environment, which should have lower metallicities
  \citep[$\sim$1/3 solar;][]{prochaska17}.  Instead, the Cloud likely
comes from M51's similarly high metallicity inner regions
\citep{bresolin04}. This in turn supports models where the Cloud is
tidally stripped gas or, alternatively, gas ejected from M51 via winds
from AGN or starburst activity. Strong outflows of gas are seen in
many starburst galaxies \citep[e.g.,][]{dahlem98} and AGN
\citep[e.g.,][]{feruglio10}, reaching large distances from the host
galaxy.

Given M51's ongoing interaction, tidal stripping is perhaps the most
obvious origin for the Cloud. This scenario is supported by the
Cloud's kinematic contiguity with M51's observed tidal \ion{H}{1}, as
well as the morphological connection between the Cloud and the NE hook
at the end of NGC~5195's western tidal tail. Further information comes
from existing numerical simulations which explore M51's interaction
history \citep[e.g.,][]{hernquist90, salo00a, durrell03}. The Cloud's
far northern location and high relative velocity provide the tightest
constraints on the models --- only the multi-passage model of
\citet{salo00a} reproduces any such gas. In this model, this gas is
first elevated above NGC~5194's disk plane in response to the
companion's initial passage, then is pulled downward along the line of
sight by the second encounter. If this model is correct, the Cloud
could represent gas initially stripped out and then met by the
companion at high velocity during the second passage, resulting in
shock-heating \citep{dopita03}.  From MAPPINGS III models
\citep{allen08}, shock velocities between $\sim$200--300 km s$^{-1}$
reproduce the Cloud's observed line ratios, in good agreement with the
  174 $\pm$ 13 km s$^{-1}$ line of sight velocity difference between
  the Cloud and NGC~5194.  However, turbulence in shocked gas
    should also broaden the lines, yet we measure typical line FWHM of
    only $\sim$60 km s$^{-1}$.  This particular model for the M51
  system also must necessarily be incomplete as it (and all other
  simulations to date) fails to reproduce all of the system's extended
  tidal features \citep{watkins15}.

While the Cloud may be tidally stripped, appealing to shock
heating as its ionization source may also suffer from a timescale
problem.  Following \citet{fossati16}, we can estimate the Cloud's
recombination timescale from its density; using an updated
  derivation of the [\ion{S}{2}] density estimator \citet{sanders16},
the Cloud's $\log$([\ion{S}{2}]$\lambda$6716$/$6731) ratio of 0.14
yields $n \sim 50$ cm$^{-3}$, for a recombination time of $\sim$2000
yr, orders of magnitude shorter than the estimated time of last
passage \citep[50--500 Myr, e.g.,][]{howard90, salo00a, durrell03}.
However, these constraints may not be so severe.  First, the
  uncertainty on the [\ion{S}{2}] line ratio encompasses the low
    density limit of the [\ion{S}{2}] density estimator, allowing for
    significantly lower densities and longer recombination times.
    Second, recently \citet{gavazzi17} argued that a diffuse cloud of
  ionized gas in the Coma Cluster could be as old as 85 Myr.
  Similarly, a network of shock-excited \ha-emitting filaments is seen
  extending $\sim$120 kpc between the galaxies M86 and NGC~4438 in the
  Virgo Cluster \citep{kenney08}; the time since closest approach
  between these two galaxies is thought to be $\sim$100 Myr ago.
  These observations suggest that some mechanism can prolong the
  lifespans of such clouds well beyond the recombination timescale.
  In galaxy clusters like Coma or Virgo, this mechanism is likely
  ongoing ram-pressure effects from the hot intracluster medium.
  While M51 is not in a cluster environment, numerical simulations
  suggest that strong galaxy interactions can lead to the formation of
  hot halo gas \citep{sinha09} and may provide a similar mechanism
  here.

Alternatively, the high line ratios observed in the Cloud could be due
to hard photoionization from either a central AGN or starburst driven
winds.  ``Hanny's Voorwerp'', a circumgalactic cloud near the fading
AGN IC~2497 \citep{keel12} provides a comparable example of the
former; in IC~2497, it is believed that a recent interaction triggered
a central AGN which illuminated and ionized the Voorwerp before
nuclear activity ceased $\sim$10$^{5}$ years ago \citep[evidenced by a
  radio jet pointing toward the Voorwerp;][]{keel12}.  A similar
scenario may be playing out in M51, where both galaxies host active
nuclei \citep{ho97, rampadarath18} and weak nuclear jets \citep[though
  only NGC~5195's jet is currently aligned with the Cloud;][]{ford85,
  rampadarath18}, although the lack of strong [\ion{O}{3}] emission
  implied by our broadband imaging suggests the Cloud could be older
  than the Voorwerp \citep{binette87}.  The currently weak nuclear
activity in M51 provides another similarity with the IC~2497 system;
following the method employed by \citet{lintott09} we find that the
current nuclear X-ray luminosity of either galaxy is too faint by four
orders of magnitude to account for the Cloud's total \hal, implying
that if the Cloud were ionized by AGN emission, the AGN has since
faded significantly.  This does not rule out past AGN activity,
however, as the strength of such activity, and the jet angle, can
evolve on $<$1 Myr timescales \citep{denney14, nawaz16}.

Finally, the Cloud may be the result of starburst driven
``superwinds'' \citep{heckman90}, in which outflows from young massive
stars and supernovae in a galaxy's disk generate expanding bubbles of
hot gas through the IGM.  These bubbles can ``blow out'' and generate
localized shocks in density inhomogeneities in the galaxy's gaseous
halo tens to hundreds of kiloparsecs from the starburst
  \citep{heckman90, heckman17}.  One nearby example is the ``cap'' of
ionized gas observed $\sim$10 kpc north of the starburst galaxy M82
\citep{devine99, lehnert99}, which lies parallel to M82's disk and in
direct line with its starburst winds \citep[clearly outlined in
  \ha\ and X-ray emission;][]{lehnert99}.  However, M51 contains no
known features indicative of superwinds, its SFR is a factor of two
lower than that in M82 \citep[after correcting for
  extinction;][]{lehnert99, kennicutt09}, and the Cloud is much more
distant from M51 than the cap is from M82 ($>$32 kpc, i.e., the
projected distance).  If the shock front is traveling at 300 km
s$^{-1}$ \citep[as we estimate from MAPPINGS III models;][]{allen08},
this implies the proposed superwind shock front has been propagating
for $>$100 Myr, significantly longer than starburst durations in most
known superwind galaxies \citep[typically of order 10$^{7}$
  yr;][]{heckman90}.  Therefore, this scenario appears unlikely unless
the Cloud's orientation with respect to NGC~5194's disk was
significantly more favorable in the past.

\section{Summary}

We report the discovery of a vast, diffuse ionized gas cloud projected
13\arcmin\ (32 kpc) north of the interacting galaxy pair M51. The
Cloud spans 13\arcmin$\times$3\arcmin\ (25$\times$7.5 kpc) in size,
and its systemic velocity (637 $\pm$ 13 km s$^{-1}$) confirms its
association with the M51 system. The Cloud has no embedded star
formation, and its high \NII/\ha, \SII/\ha, and \OI/\ha\ line ratios
suggest AGN photoionization or shock heating.  While not directly
overlapping with M51's tidal features, the Cloud shows mild evidence
of morphological contiguity with the companion galaxy's bifurcated
western arm, and kinematic similarity to adjacent tidally stripped \ion{H}{1}.

The strong \NII\ emission implies high solar-like metallicities, such
that rather than being primordial infall, the Cloud has likely been
expelled from the inner disk of NGC~5194 via tidal stripping or
AGN/starburst winds.  The currently low level of nuclear activity in
both galaxies implies that if the Cloud were ionized by AGN activity,
this activity has since faded \citep[a situation similar to IC~2497
  and Hanny's Voorwerp;][]{lintott09, keel12}.  Alternatively, the
Cloud may have been ejected by starburst driven ``superwinds''
\citep{heckman90}, however M51's low SFR relative to known superwind
galaxies \citep[e.g., M82;][]{devine99, lehnert99}, as well as the
Cloud's extreme distance from M51 ($>$32 kpc), suggests that this
scenario is less likely.  Finally, the gas may be shock heated
due to the ongoing interaction between the galaxy pair; if so, it may
lend support to the multi-passage interaction model of M51 proposed by
\citet{salo00a}, to date the only such model to produce significant
high velocity gas north of the system.

To discriminate between these various scenarios for the Cloud's
origin, additional spectroscopic observations are needed which target
emission lines that probe the Cloud's density and temperature
structure, better constrain its metallicity, and differentiate between
photoionization and shock-heating models for the Cloud. Additional
information would come from mapping the line ratios and kinematics of
the Cloud across its spatial extent as well.

The Cloud's size and structure --- and, perhaps most importantly, the
proximity of the M51 system --- provide a unique opportunity to study
the detailed effects of feedback and ionization on the circumgalactic
environments of galaxies. The local universe contains very few known
examples of extended diffuse emission sources like the Cloud; each new
example provides a wealth of new information about tidal interactions,
feedback processes, and the mutual interaction between galaxies and
their environment. In particular, if the Cloud is a fossil nebula
  or echo of strong AGN activity in M51, it would be the most nearby
example of a rapidly fading AGN, and also represent a new and critical
piece to our understanding of the iconic M51 system.

\acknowledgements
We thank Heikki Salo and Eija Laurikainen for useful discussions
regarding dynamical models of M51, and Tim Heckman, Bill
  Keel, and the anonymous referee for useful suggestions. Support for
this project was provided by NSF/AST-1108964 (JCM) and NSF/AST-1517006
(MAB).

{\it Facilities:}
\facility{CWRU:Schmidt},  \facility{WIYN}

\end{document}